\documentclass[aps,prl,twocolumn,showpacs,preprintnumbers,amsmath,amssymb,floatfix,10pt]{revtex4}
\usepackage{times}
\usepackage{graphicx}
\usepackage{dcolumn}
\usepackage{bm}
\usepackage{subfigure}
\usepackage{mathrsfs}
\usepackage{bbm}
\usepackage{textcomp}

\begin{document}
\title{Quantum radiation reaction effects in multiphoton Compton scattering}

\author{A. Di Piazza}
\email{dipiazza@mpi-hd.mpg.de}

\author{K. Z. Hatsagortsyan}

\author{C. H. Keitel}
\affiliation{Max-Planck-Institut f\"ur Kernphysik, Saupfercheckweg 1, 69117 Heidelberg, Germany}

\date{\today}
\begin{abstract}
Radiation reaction effects in the interaction of an electron and a strong laser field are investigated in the realm of quantum electrodynamics. We identify quantum radiation reaction with the multiple photon recoils experienced by the laser-driven electron due to consecutive incoherent photon emissions. After determining a quantum radiation dominated regime, we demonstrate how in this regime quantum signatures of radiation reaction strongly affect multiphoton Compton scattering spectra and that they could be measurable in principle with presently available laser technology.

\end{abstract}

\pacs{12.20.Ds, 41.60.-m}
\maketitle

Radiation reaction (RR) represents one of the oldest and still open problems in electrodynamics \cite{Landau_b_2_1975,Books_RR}. In classical electrodynamics a charged particle in the presence of an external electromagnetic field emits electromagnetic radiation, continuously looses energy and momentum so that the emission modifies the trajectory of the particle itself. To take into account this ``reaction'' of the emitted radiation on the electron motion one should solve Lorentz equation together with Maxwell equations self-consistently. This amounts in solving the so-called Lorentz-Abraham-Dirac (LAD) equation of motion \cite{Books_RR}. This equation, although solving in principle the problem of RR classically, is plagued by inconsistencies, like the existence of the so-called runaway solutions \cite{Landau_b_2_1975,Books_RR}. However, these inconsistencies are fictitious in the realm of classical electrodynamics as they arise at length and time scales well in the quantum regime \cite{Landau_b_2_1975}. Moreover, in classical electrodynamics the LAD equation can be consistently approximated by the so-called Landau-Lifshitz (LL) equation, which does not show the shortcomings of the LAD equation \cite{Landau_b_2_1975,Spohn_2000,Rohrlich_2002}. The LL equation has not yet been tested experimentally and proposals have been put forward to achieve this goal \cite{Koga_2005,Di_Piazza_2009}. An alternative classical equation of motion including the RR phenomenologically has been proposed in Ref. \cite{Sokolov_2009}.

Though in the realm of classical electrodynamics the force on the electron due to RR is much smaller than the Lorentz force in the instantaneous rest frame of the electron, it can be the dominant force in the laboratory frame \cite{Landau_b_2_1975}. Accordingly, the so-called (classical) radiation dominated regime (RDR) for Thomson scattering can be identified \cite{Koga_2005,Di_Piazza_2008}. In the case of an external laser field, the RDR is entered when the energy radiated during one driving laser period is of the order of or exceeds the initial electron energy. For a plane wave laser field with carrier angular frequency $\omega_0$ and electric field amplitude $E_0$, and for an electron (mass $m$ and charge $-e<0$), initially counterpropagating with the laser field, with initial energy $\varepsilon_0$ and momentum $p_0>0$, the average energy emitted by the electron per unit time is of the order of $\alpha m^2\chi_0^{\,2}$, where $\alpha=e^2\approx 1/137$, $\chi_0=(\varepsilon_0+p_0)E_0/mE_{cr}$, with $E_{cr}=m^2/e=1.3\times 10^{16}\;\text{V/cm}$ being the critical field of quantum electrodynamics (QED) (units with $\hbar=c=1$ are used throughout) \cite{Ritus_Review_1985}. Accordingly, the classical RDR is characterized by the condition \cite{Koga_2005,Di_Piazza_2008}: $R_C\equiv\alpha \chi_0 \xi_0 \gtrsim 1$, where $\xi_0=eE_0/m\omega_0$ is the relativistic field parameter, assumed here to be much larger than unity (present laser systems already allow for values of $\xi_0\sim 100$ \cite{Yanovsky_2008}).

The above considerations are classical. Quantum effects become important if $\chi_0\gtrsim 1$. In fact, the motion of an ultrarelativistic electron in a laser field is quasiclassical \cite{Baier_b_1994}. Quantum effects amount to the recoil experienced by the electron in the photon emission which is of the order of $\chi_0\varepsilon_0$ \cite{Ritus_Review_1985}. However, quantum RR effects are not exhausted by single-photon recoil. In fact, quantum single-photon emission spectra at $\chi_0\ll 1$ turn into the corresponding classical spectra but without RR effects included.

In the present Letter we investigate the quantum RDR in the interaction between an electron and an intense laser field. We identify quantum RR with the consecutive photon recoils in multiple incoherent single-photon emissions by the laser-driven electron, i.e. in the successive single-photon emissions each occurring in different radiation formation lengths. The quantum RDR is characterized by the two parametric conditions: $\chi_0\gtrsim 1$ and $R_Q\equiv\alpha \xi_0\gtrsim 1$. The first condition implies that the recoil in each photon emission is in general significant and it is fully accounted for by working in the framework of strong-field QED \cite{Ritus_Review_1985}. The second condition, instead, involves the quantum parameter $R_Q$ and implies that the average number of photons emitted incoherently in one laser period can be larger than unity. We employ a new, microscopic approach to describe multiple incoherent photoemissions, alternative to the usual kinetic one \cite{Baier_b_1994,Baier_1999,Khokonov_2004}, which allows us to obtain emission spectra without solving the kinetic partial integro-differential equations. In this approach, the radiation process consists of multiple channels, each corresponding to a different number of incoherently emitted photons. The change of the electron state is taken into account consistently at each emission event which happens in a statistically uncorrelated way. This is in contrast to the approach in \cite{Sokolov_2010} where, the statistical character of the photon emission is neglected. We calculate numerically quantum emission spectra with and without RR and show how significantly the photon spectrum due to multiple emissions may differ from that predicted in the case of a single emission. This can represent a possible observable to measure RR effects in the quantum regime at optical laser intensities of the order of $10^{22}$ W/cm$^2$ \cite{Yanovsky_2008}.

An ultrarelativistic electron with initial four-momentum $p_0^{\,\mu}=(\varepsilon_0,0,-p_0,0)$ ($\varepsilon_0=\sqrt{m^2+p_0^2}$ and $p_0>0$) is considered to collide head-on with a plane wave characterized by the four-vector potential $A_0^{\,\mu}(\phi)=(0,\hat{\bm{z}}A_0f(\phi))$, where $\phi=\omega_0(t-y)$, $A_0=-E_0/\omega_0$ and $f(\phi)$ is 
an arbitrary function with $|f'(\phi)|_{\text{max}}= 1$ ($f'(\phi)=df(\phi)/d\phi$). The external plane wave field can be taken into account exactly by quantizing the Dirac field in the Furry picture and by employing so-called Volkov states as electron states [10]. On the other hand, the interaction between the Dirac field and the photon field is taken into account perturbatively up to first order, i.e. in each formation length only the emission of one photon is taken into account. If $\chi_0\lesssim 1$ [$\chi_0\gg 1$] the emission probability of $j$ photons in the same radiation formation length is approximately $\alpha^{j-1}$ [$(\alpha\chi_0^{\,2/3})^{j-1}$] times the probability of emission of one photon [10], therefore we restrict ourselves here to $\chi_0\ll 1/\alpha^{3/2}\sim 10^3$ [16]. In general, it is convenient to label the particle state with initial four-momentum $p^{\mu}=(\varepsilon,\bm{p})=(\sqrt{m^2+\bm{p}^2},\bm{p})$ via the quantity $p_-\equiv\varepsilon-p_y$, which is a constant of motion in the presence of the plane wave with four-vector potential $A_0^{\,\mu}(\phi)$. Also, we will always consider situations in which $\varepsilon_0\gg m\xi_0$, therefore the transverse degrees of freedom of the particles in the problem can be neglected and only probabilities integrated over the transverse momenta will be considered. We also consider probabilities averaged over the initial electron spin and summed over the final electron spin and the emitted photon polarization. As a starting point, we employ the expression for the probability of one photon emission $dP_{p_-}^{(1)}/dud\phi$ by an electron with initial four-momentum $p^{\mu}$ in the external plane wave per unit of the laser phase $\phi$ and per unit $u=k_-/(p_--k_-)$, with $u\in [0,\infty)$ \cite{Ritus_Review_1985}:
\begin{equation}
\label{dP_1/dudphi}
\begin{split}
\frac{dP_{p_-}^{(1)}}{dud\phi}&=\frac{\alpha}{\pi\sqrt{3}}\frac{m^2}{\omega_0p_-}\frac{1}{(1+u)^2}\Bigg[\left(1+u+\frac{1}{1+u}\right)\\
&\left.\times K_{2/3}\left(\frac{2u}{3\chi(\phi)}\right)-\int_{2u/3\chi(\phi)}^{\infty}dyK_{1/3}(y)\right],
\end{split}
\end{equation}
where $k_-=\omega-k_y$, with $k^{\mu}=(\omega,\bm{k})=(|\bm{k}|,\bm{k})$ being the four-momentum of the emitted photon, where $K_{\nu}(x)$ is the modified Bessel function of $\nu$th order and where $\chi(\phi)=p_-|E(\phi)|/mE_{cr}$, with $E(\phi)=E_0f'(\phi)$ being the instantaneous electric field of the plane wave. Note that $\chi(\phi)=(p_-/p_{0,-})\chi_0|f'(\phi)|=(p_-/p_{0,-})\chi_0(\phi)$, with $\chi_0(\phi)=\chi_0|f'(\phi)|$. The above expression of $dP_{p_-}^{(1)}/dud\phi$ is valid in the limit of a constant crossed field when $\xi_0\gg 1$. However, since in this limit the radiation formation length of the photon production process is of the order of $\lambda_0/\xi_0\ll \lambda_0$, with $\lambda_0=2\pi/\omega_0$ being the central wavelength of the plane wave \cite{Ritus_Review_1985}, then the same expression of the probability can be employed for an external laser field varying in a space scale of the order of $\lambda_0$, but with the instantaneous value of the electric field strength. In this respect, the instantaneous value $p_-(\phi)$ at the moment of emission should be employed in Eq. (\ref{dP_1/dudphi}), but in a plane wave the quantity $p_-(\phi)$ is constant: $p_-(\phi)\equiv p_-$.

For sufficiently long and/or strong pulses the quantity $P_{p_{0,-}}^{(1)}=\int_{-\infty}^{\infty}d\phi\int_0^{\infty} du\, dP_{p_{0,-}}^{(1)}/dud\phi$ becomes larger than unity, and it cannot be interpreted as an emission probability. In this case multiple incoherent photon emissions take place, i.e. the electron emits many photons but each in a different radiation formation length. In the classical limit of emission of low-energy photons with respect to the initial energy of the electron, this apparent contradiction can be easily resolved following the procedure developed for the so-called ``infrared catastrophe'' (see \cite{Infrared_Catastrophe}). The conclusion is that in the classical limit the quantity $P_{p_{0,-}}^{(1)}$ is actually the average number of photons emitted (see also \cite{Khokonov_2004}) and, therefore, it is not contradictory that it can be larger than unity \cite{Infrared_Catastrophe}. In the quantum regime we proceed analogously by taking into account, however, that at each photon emission step the electron incoming quantum numbers are changed and determined by the previous step (see Fig. 1).
\begin{figure}
\begin{center}
\includegraphics[width=7cm]{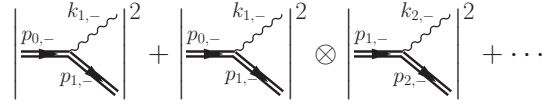}
\caption{A diagrammatic visualization of the calculation of incoherent multiphoton emission by a laser-driven electron. The double electron lines indicate that the electron states are Volkov states and the dots represent the emission of more than two incoherent photons. At each photon emission step the electron incoming quantum number $p_-$ is changed and determined by the previous step.}
\end{center}
\end{figure}
The total probability $P_{p_{0,-}}^{(N)}$ that an electron with initial momentum $p_0^{\mu}$ emits $N$ photons incoherently is given by $P_{p_{0,-}}^{(N)}=\int [du]_N \int[d\phi]_NdP_{p_{N-1,-}}^{(1)}/du_Nd\phi_N\cdots dP_{p_{0,-}}^{(1)}/du_1d\phi_1$, where $p_{j,-}=p_{j-1,-}/(1+u_j)$ with $j=1,\ldots, N-1$ and where the notation $\int [du]_N\equiv\int_0^{\infty}du_N\int_0^{\infty}du_{N-1}\cdots\int_0^{\infty}du_1$ and $\int [d\phi]_N\equiv\int_{-\infty}^{\infty}d\phi_N\int_{-\infty}^{\phi_N}d\phi_{N-1}\cdots\int_{-\infty}^{\phi_2}d\phi_1$ has been introduced. Since Volkov states are normalized to unity, the probability of no-photon emission is also unity at this stage. However, since the total probability that either no-photons or any number of photons are emitted must be unity, the actual total probability $\mathcal{P}_{p_{0,-}}^{(N)}$ of emission of $N$ photons should be renormalized as $\mathcal{P}_{p_{0,-}}^{(N)}=P_{p_{0,-}}^{(N)}/\mathcal{N}_{p_{0,-}}$, with $\mathcal{N}_{p_{0,-}}=1+P_{p_{0,-}}^{(1)}+\cdots+P_{p_{0,-}}^{(N)}+\cdots$. Analogously, the probability of no-photon emission is actually $1/\mathcal{N}_{p_{0,-}}$. In the classical limit, i.e. when $\chi_0\ll 1$, then the main contribution to the integral in $P_{p_{0,-}}^{(N)}$ comes from the region $u_j\ll 1$ for $j=1,\ldots, N$, then the multidimensional integral in $P_{p_{0,-}}^{(N)}$ becomes factorisable and we obtain the Bloch and Nordsieck result: $P_{p_{0,-}}^{(N)}=(P_{p_{0,-}}^{(1)})^N/N!$ and $\mathcal{N}_{p_{0,-}}=\exp(P_{p_{0,-}}^{(1)})$ \cite{Infrared_Catastrophe}. In contrast to that, in the quantum regime the individual probabilities $dP_{p_{j-1,-}}^{(1)}/du_jd\phi_j$ in $P_{p_{0,-}}^{(N)}$ are not independent from each other, because, as we have mentioned, the initial electron momentum at each emission is different. The effects of RR arise here from the fact that the emission of each photon modifies the electron state and consequently the next emissions. Note that the quasiclassical motion of the electron between two emissions is taken into account consistently here, in the sense that the only dynamical quantity of the electron appearing in the equations, i.e. $p_{j,-}$, does not change between two emissions. Now, we introduce the quantum photon spectrum $d\mathcal{S}_Q^{\text{RR}}/d\varpi$ including RR, of a photon with $k_-=\varpi p_{0,-}$ emitted by an electron with initial $p_{0,-}$ (for notational simplicity the index $p_{0,-}$ is not included in $d\mathcal{S}_Q^{\text{RR}}/d\varpi$):
\begin{equation}
\label{Spectrum_Q_RR}
\begin{split}
\frac{d\mathcal{S}_Q^{\text{RR}}}{d\varpi}=&\frac{\varpi}{\mathcal{N}_{p_{0,-}}}\bigg[\int[d\phi]_1\frac{dP_{p_{0,-}}^{(1)}}{d\varpi d\phi_1}+\sum_{N=2}^{\infty}\int[du]_{N-1}\\
&\times\int [d\phi]_N\sum_{i=1}^N\vartheta\bigg(\prod_{l=0}^{i-1}\frac{1}{1+u_l}-\varpi\bigg)\\
&\times\frac{dP_{\tilde{p}_{N-1,-}}^{(1)}}{du_{N-1} d\phi_N}\cdots\frac{dP_{\tilde{p}_{i,-}}^{(1)}}{du_i d\phi_{i+1}}\frac{dP_{\tilde{p}_{i-1,-}}^{(1)}}{d\varpi d\phi_i}\\
&\times\frac{dP_{\tilde{p}_{i-2,-}}^{(1)}}{du_{i-1} d\phi_{i-1}}\cdots\frac{dP_{p_{0,-}}^{(1)}}{du_1 d\phi_1}\bigg].
\end{split}
\end{equation}
In this expression the $i$th term in the sum over $i$ corresponds to the photon with $k_-$ being emitted at $\phi_i$. In general, $dP_{p_-}^{(1)}/d\varpi d\phi=(p_-/p_{0,-})(p_-/p_{0,-}-\varpi)^{-2}dP_{p_-}^{(1)}/dud\phi$ and $\varpi=(p_-/p_{0,-})u/(1+u)< 1$, while the step function $\vartheta(x)$ ensures that the corresponding term vanishes if the electron has not enough energy to emit the photon with $k_-$ ($u_0\equiv 0$). Furthermore, $\tilde{p}_{j,-}=\tilde{p}_{j-1,-}-\varpi p_{0,-}$ [$\tilde{p}_{j,-}=\tilde{p}_{j-1,-}/(1+u_j)$] if the $j$th photon emitted is the on with $k_-=\varpi p_{0,-}$ [$k_{j,-}=\tilde{p}_{j-1,-}u_j/(1+u_j)$] for $j=1,\ldots, N-1$ ($\tilde{p}_{0,-}=p_{0,-}$). Since we assumed that $\varepsilon_0\gg m\xi_0$ then $p_{0,-}=\varepsilon_0-p_{0,y}\approx 2\varepsilon_0$ and the photons are mostly emitted along the negative $y$-direction, i.e. $k_-=\omega-k_y\approx 2\omega$, then the quantity $d\mathcal{S}_Q^{\text{RR}}/d\varpi$ is essentially the differential average energy emitted in units of the initial electron energy. From the above discussion it is clear that in order to be in the quantum RR regime it must be: $P_{p_{0,-}}^{(1)}\gtrsim 1$ (significant incoherent emission of many photons during the whole laser pulse) and $\chi_0\gtrsim 1$ (non-negligible photon recoil). Starting from Eq. (\ref{dP_1/dudphi}) with $p_-=p_{0,-}$ and by performing the variable change $u'=2u/3\chi_0(\phi)$ in $P_{p_{0,-}}^{(1)}$, one sees that if $\chi_0\lesssim 1$, then $P_{p_{0,-}}^{(1)}\sim \alpha(m^2/\omega_0 p_{0,-})\chi_0\Delta \phi=\alpha\xi_0\Delta \phi$, where $\Delta \phi$ is the phase interval corresponding to the pulse duration. Correspondingly, the analogous condition of being in the quantum RDR, i.e. $R_Q=\alpha\xi_0\gtrsim 1$, is obtained from the above one by setting $\Delta \phi\sim 1$ (one period corresponds actually to $\Delta \phi=2\pi$). Therefore, when RR is not important, i.e. when $\alpha \xi_0 \Delta \phi \ll 1$, the emitted spectrum is given by the quantity $d\mathcal{S}_Q^{\text{no RR}}/d\varpi$ defined as $d\mathcal{S}_Q^{\text{no RR}}/d\varpi\equiv\varpi dP_{p_{0,-}}^{(1)}/d\varpi\equiv\varpi\int [d\phi]_1 dP_{p_{0,-}}^{(1)}/d\varpi d\phi_1$.

Below we compare the quantum spectra $d\mathcal{S}_Q^{\text{RR}}/d\varpi$ and $d\mathcal{S}_Q^{\text{no RR}}/d\varpi$ also with the corresponding classical quantities. The expressions $d\mathcal{S}_C^{\text{no RR}}/d\varpi$ and $d\mathcal{S}_C^{\text{RR}}/d\varpi$ of the classical spectrum without and with RR can be obtained starting from Eq. (\ref{dP_1/dudphi}) and from $d\mathcal{S}_Q^{\text{no RR}}/d\varpi$ for an electron with initial momentum $p_0^{\,\mu}$ in the limit $k_-\ll p_{0,-}$ (see also \cite{Baier_b_1994}). While without RR it is $p_-(\phi)=p_{0,-}\equiv p_-^{\text{no RR}}(\phi)$, the analytical solution of the LL equation found in \cite{Di_Piazza_2008} indicates that if RR is included, it is $p_-(\phi)=p_{0,-}/h(\phi)\equiv p_-^{\text{RR}}(\phi)$, with $h(\phi)=1+(2/3)R_C\int_{-\infty}^{\phi}d\phi'f^{\prime\, 2}(\phi')$. Therefore, we have $d\mathcal{S}_C^{\text{RR/no RR}}/d\varpi\equiv\int [d\phi]_1d\mathcal{S}_C^{\text{RR/no RR}}/d\varpi d\phi_1$, with
\begin{equation}
\frac{d\mathcal{S}_C^{\text{RR/no RR}}}{d\varpi d\phi}\equiv\varpi\left.\frac{dP_{p_{0,-}}^{(1)}}{d\varpi d\phi}\right\vert_{\begin{subarray}{l}p_-=p_-^{\text{RR/no RR}}(\phi)\\ \varpi\ll 1\end{subarray}}.
\end{equation}
%
If one calculates classically the average energy $\langle \omega\rangle$ emitted in one laser period as $\langle \omega\rangle\approx\varepsilon_0(2\pi)^{-1}\int_0^{2\pi}d\phi\int_0^1 d\varpi d\mathcal{S}_C^{\text{no RR}}/d\varpi d\phi$, one obtains that the condition $\langle \omega\rangle\gtrsim \varepsilon_0$ is fulfilled when $R_C\gtrsim 1$, which is the classical RDR condition. In the classical regime, accounting for RR amounts essentially to subtract at each instant of time $t+dt$, from the four-momentum of the electron at time $t$ the \emph{average} four-momentum that it emits as radiation in the short interval $dt$. In contrast to that, in the quantum regime the loss of energy and momentum by the electron through the emission of a photon is intrinsically a probabilistic event. Only if the emission occurs, then the state of the electron and the successive emissions are modified. If one calculates $\langle \omega\rangle$ starting from $d\mathcal{S}_Q^{\text{no RR}}/d\varpi$, one sees that quantum corrections are negative \cite{Ritus_Review_1985}, thus the classical ``averaged'' treatment of RR is expected to overestimate the effects of RR when quantum effects become important.

Below, we shall provide a numerical example in order to show that in principle the quantum RR regime can be already experimentally realized with presently available laser technology. We consider the simple case of a two-cycle sinusoidal pulse with $f(\phi)=-\cos(\phi)$ and $\phi\in[0,4\pi]$. In this case the integrations in $\phi$ in Eq. (\ref{Spectrum_Q_RR}) can be approximately performed analytically and the numerical calculation of the remaining integrals via the Monte Carlo method becomes feasible (numerical time-consuming Bessel functions can be avoided by employing an alternative equivalent representation of $dP_{p_{0,-}}^{(1)}/dud\phi$ as in Eq. (4.2) in \cite{Khokonov_2004}).
\begin{figure}
\begin{center}
\includegraphics[width=7cm]{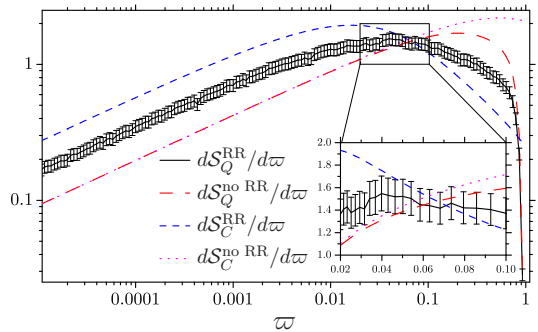}
\caption{(Color online) Multiphoton Compton spectra calculated quantum mechanically with (solid, black line) and without (long dashed, red line) RR. For the sake of comparison, the corresponding classical spectra with (short dashed, blue line) and without (dotted, magenta line) RR are also shown. The error bars in $d\mathcal{S}_Q^{\text{RR}}/d\varpi$ stem from uncertainties in the multidimensional integrations in Eq. (\ref{Spectrum_Q_RR}). The inset shows a zoom of the spectral region where different curves cross. The numerical parameters are: $\varepsilon_0=1\;\text{GeV}$, $\omega_0=1.55\;\text{eV}$ and $I_0=E_0^{\,2}/8\pi=5\times 10^{22}\;\text{W/cm$^2$}$, corresponding to $\xi_0=154$, $\chi_0=1.8$ and $R_Q=1.1$.}
\end{center}
\end{figure}
Quantum and classical spectra with and without RR shown in Fig. 2 clearly display different behaviors. Since classically the emitted frequencies do not have the meaning of photon energies, the classical spectra allow for the unphysical emission at $\varpi\ge 1$, also in the case that RR is taken into account. This is forbidden quantum mechanically and the existence of the physical boundary $\varpi=1$ originates a typical quantum piling-up of the radiation by an ultra-relativistic electron towards $\varpi=1$. Inclusion of RR in the quantum regime has mainly three effects: 1) increase of the spectral yield at low energies, 2) shift to lower energies of the maximum of the spectral yield and 3) decrease of the spectral yield at high energies. The physical reason is that, due to RR, the electron looses its energy by emitting several relatively low-energy photons and the probability of emitting one photon in the region very close to $\varpi\approx 1$ is less than if RR is neglected. Fig. 2 also shows that the classical treatment of RR (short dashed, blue curve) artificially enhances the above three effects of RR and, as we have mentioned, this is mainly due to the classical overestimation of the average energy emitted by the electron. Finally, in Fig. 2 emission of up to sixteen photons is taken into account implying the numerical integration of up to $15^{th}$-fold integrals, the remaining higher-order terms having been estimated to give a contribution of about $2\;\%$.

Here we have not considered the exact space-time shape of the laser field. It is however clear from the above physical considerations, that the laser's pulse shape will not change qualitatively our predictions. Concerning the observability of the discussed effect, one could expect that the emitted photons, by interacting again with the laser field might create an electron-positron pair and start a cascade process \cite{Fedotov_2009}. However, from Fig. 2 we see that the photon spectrum $d\mathcal{S}_Q^{\text{RR}}/d\varpi$ has a peak at $\varpi_0\approx 0.04$. The probability of pair creation by a created photon with a given $\varpi$ is roughly suppressed by a factor $\eta(\varpi)=\exp(-8/3\chi_{\gamma}(\varpi))$, with $\chi_{\gamma}(\varpi)=\varpi\chi_0$ assumed to be much smaller than unity \cite{Ritus_Review_1985}. Since in the numerical example $\chi_0=1.8$, it is $\chi_{\gamma}(\varpi_0)=0.073$ and $\eta(\varpi_0)\sim 10^{-16}$. Thus, we can conclude that most of the created photons, at not too large $\varpi$'s, will escape the laser pulse and reach the detector. Finally, we have calculated from Fig. 2 that about $1.74\pm 0.17$ and $1.00$ photons are expected to be emitted at $10^{-5}<\varpi<10^{-3}$ (corresponding to photon energies between $10\;\text{KeV}$ and $1\;\text{MeV}$) per electron with and without RR, respectively, amounting to a relative difference of about $(43\pm 11)\;\%$ (note that for this example substantial deviations from Eq. (\ref{dP_1/dudphi}) according to \cite{Khokonov_2002_b} and consequently from the estimated photon yields only apply at $\varpi<10^{-8}$). This and the fact that laser intensities and electron energies considered above are within the reach of present technology, allow in principle for experimental measurement of the quantum RR effects with an all-optical experiment exploiting electron wake-field acceleration \cite{Yanovsky_2008,Leemans_2006}.


\begin{thebibliography}{39}
\bibitem{Landau_b_2_1975} L. D. Landau, and E. M. Lifshitz, \textit{The Classical Theory of Fields}, (Elsevier, Oxford, 1975).

\bibitem{Books_RR} H. Spohn, \textit{Dynamics of charged particles and their radiation field}, (Cambridge University Press, Cambridge, 2004); F. Rohrlich, \textit{Classical Charged Particles}, (World Scientific, Singapore, 2007).


\bibitem{Spohn_2000} H. Spohn, Europhys. Lett. \textbf{50}, 287 (2000).

\bibitem{Rohrlich_2002} F. Rohrlich, Phys. Lett. A \textbf{303}, 307 (2002).

\bibitem{Koga_2005} J. Koga et al., Phys. Plasmas \textbf{12}, 093106 (2005).

\bibitem{Di_Piazza_2009} A. Di Piazza, K. Z. Hatsagortsyan, and C. H. Keitel, Phys. Rev. Lett. \textbf{102}, 254802 (2009).

\bibitem{Sokolov_2009} I. V. Sokolov et al., Phys. Plasmas \textbf{16}, 093115 (2009).

\bibitem{Di_Piazza_2008} A. Di Piazza, Lett. Math. Phys. \textbf{83}, 305 (2008).

\bibitem{Ritus_Review_1985} V. I. Ritus, J. Sov. Laser Res. \textbf{6}, 497 (1985).

\bibitem{Yanovsky_2008} V. Yanovsky et al., Opt. Express {\bf 16}, 2109 (2008).

\bibitem{Baier_b_1994} V. N. Baier, V. M. Katkov, and V. M. Strakhovenko, \textit{Electromagnetic Processes at High Energies in Oriented Single Crystals}, (World Scientific, Singapore, 1998).

\bibitem{Baier_1999} V. N. Baier and V. M. Katkov, Phys. Rev. D \textbf{59}, 056003 (1999).

\bibitem{Khokonov_2004} M. Kh. Khokonov, Soviet Phys. JETP \textbf{99}, 690 (2004).
\bibitem{Sokolov_2010} I. V. Sokolov et al., Phys. Rev. E \textbf{81}, 036412 (2010).

\bibitem{Footnote_2} If $\alpha\chi_0^{\,2/3}\gtrsim 1$ it also happens that radiative corrections become of the same order of the tree-level results \cite{Ritus_Review_1985} and the emission of radiation may eventually affect the radiation formation length itself.

\bibitem{Infrared_Catastrophe} F. Bloch and A. Nordsieck, Phys. Rev. \textbf{52}, 929 (1948); R. J. Glauber, Phys. Rev. \textbf{84}, 395 (1951).


\bibitem{Fedotov_2009} A. R. Bell and J. G. Kirk, Phys. Rev. Lett. \textbf{101}, 200403 (2008); A. M. Fedotov \textit{et al.}, Phys. Rev. Lett. \textbf{105}, 080402 (2010).

\bibitem{Khokonov_2002_b} M. Kh. Khokonov and H. Nitta, Phys. Rev. Lett. \textbf{89}, 094801 (2002).
\bibitem{Leemans_2006} W. P. Leemans et al., Nature Phys. \textbf{2}, 696 (2006).

\end{thebibliography}
\end{document}